\documentclass[11pt, letterpaper]{article}
\usepackage[top=0.85in,footskip=0.75in]{geometry}

\usepackage{lmodern}
\usepackage{amssymb,amsmath}
\usepackage[T1]{fontenc}
\usepackage[utf8]{inputenc}
\usepackage{microtype}
\UseMicrotypeSet[protrusion]{basicmath} %

\PassOptionsToPackage{hyphens}{url} %
\usepackage[unicode=true]{hyperref}
\hypersetup{pdfborder={0 0 0}, breaklinks=true, colorlinks=true, allcolors=blue}
\usepackage{parskip}
\makeatletter
\def\fps@figure{htbp}
\makeatother

\usepackage{longtable, booktabs}
\usepackage{graphicx}

\usepackage{subfig}  %

\usepackage[onehalfspacing]{setspace}
\usepackage[style=authoryear]{biblatex}
\DeclareMathOperator{\E}{E}
\DeclareMathOperator{\Var}{Var}
\DeclareMathOperator{\logit}{logit}

\title{Reassembling the English Novel, 1789-1919}
\author{
  Allen Riddell\thanks{Indiana University, Bloomington} \and
  Michael Betancourt\thanks{Symplectomorphic, LLC. Research conducted while at Columbia University.}
}

\date{2. November 2020}  %

\begin{document}

\maketitle
\newcommand{\NewNovelsSeventeenHundredEightyNineToNineteenHundredNineteenFifthPercentile}{40,000}  %
\newcommand{\NewNovelsSeventeenHundredEightyNineToNineteenHundredNineteenNinetyFifthPercentile}{63,000}  %

\newcommand{\NewNovelsEighteenHundredToEighteenHundredNinetyNineFifthPercentile}{21,000}  %
\newcommand{\NewNovelsEighteenHundredToEighteenHundredNinetyNineNinetyFifthPercentile}{28,000}  %

\newcommand{\numNewNovelsPerMillionReadersEighteenTwenty}{6.0}
\newcommand{\numNewNovelsPerMillionReadersEighteenFifty}{10.5}
\newcommand{\periodOfGrowthStart}{1840}
\newcommand{\periodOfGrowthEnd}{1855}

\section*{Introduction}

Aspirations in the 20th century for sociologically-inclined literary history foundered due to a lack
of accessible, trustworthy, and inclusive bibliographies and biographical records. Despite sustained
interest, no principled estimates of the number of novelists writing or the number of new novels
published during the 19th and early 20th centuries ever materialized
\parencite{sutherland1988publishing}. Without a detailed accounting of novelistic production,
numerous questions proved impossible to answer. The following three are representative: How many
writers made careers as novelists, Are there unacknowledged precursors or forgotten rivals to
canonical authors, To what extent is a writer's critical or commercial success predictable from
their social origins? Although material traces of every novel published in Europe and North America
survive, gathering particulars required to answer questions such as these proved too time-consuming
or too resource intensive.

The lack of credible information about the population of novelists and the
population of published novels obstructs research in literary studies, cultural studies, book
history, and sociology of literature. Two communities in particular stand to gain from a more
detailed accounting of these two populations. The first includes those interested in studying
literary form and prose style from below. A characteristic concern of this group is an interest in
how the emergence and diffusion of literary morphology reveals information about broader economic,
social, and cultural relationships within and across national and linguistic situations (e.g.,
\textcites{escarpit1958sociologie}{moretti1998atlas}{casanova1999republique}{moretti2000conjectures}).
The second group includes researchers in cultural studies and sociology of culture interested in
uniting literary history with sociological concerns.  This group includes those interested in the
working conditions facing novelists and those studying the history of occupational gender
segregation in the text industry (e.g., \textcite{williams1965long}, \textcite{tuchman1989edging}).
This group also includes those interested in reassembling an understanding of literary artworks as
products of networks of actors whose actions are necessary for works' existence and whose actions,
in turn, shape the art objects \parencite[xii]{becker1995introduction}.  Library digitization and
sharing of machine-readable datasets are two developments which support research agendas associated
with these communities. More generally, these developments facilitate studying literary works at
multiple scales and with a broader range of vocabularies.

To demonstrate the improving prospects for data-intensive, sociologically-inclined literary history---enabled by
the availability of digital surrogates of surviving volumes and the sharing of machine-readable bibliographic data---
this paper estimates the yearly rates of new novel publication in the British
Isles and Ireland between 1789 and 1919. This period witnessed, in aggregate, the publication of between
\NewNovelsSeventeenHundredEightyNineToNineteenHundredNineteenFifthPercentile{} and
\NewNovelsSeventeenHundredEightyNineToNineteenHundredNineteenNinetyFifthPercentile{} previously
unpublished novels (``new novels''). Although there has been considerable speculation about this
time series, ours are the first principled estimates to be published. The years studied include the
rise of mass literacy and one of the more important periods in the history of publishing
(1830-1850), a period during which practices and institutional arrangements resembling the modern
publishing industry emerge \parencite[328-329]{raven2007business}.

The analysis presented here is limited to literary production on islands in the North Atlantic.
Although the prospect of comparative research was a primary motivation for this work, a lack of
comprehensive bibliographical records outside the British Isles and Ireland made such research difficult.  The
exhaustive bibliography of novels published between 1770 and 1836 found in
\textcites{raven2000english}{garside2000english} (hereafter ``RFGS'')---indispensable to the work
here---has no real equivalent.  (For example, although \textcite{brummer1884lexikon} is impressive
in the number of German-language titles it documents, like \textcite{block1961english}, it makes no
claims to have enumerated all titles published.) Bibliographic work on novels written in languages
other than English is, however, ongoing and library digitization makes the work easier.
And the estimates presented here provide information about plausible trajectories of literary
production elsewhere. For example, because it is hard to imagine per capita novelistic production
growing considerably faster than it did in the British Isles and Ireland during the 1840s, the pace of growth
during this decade may be used as an estimate of the upper bound on the pace of growth in
established text industries in other regions.

\section*{Rise of the Text Industry}

No comprehensive survey of new novels published in the British Isles and Ireland exists for any year after 1836. There is
neither an exhaustive list of new novels published nor principled estimates of the number of new
novels published in any year after 1836. Given the pace of expansion in the publishing industry
during the period and the time and resources required to complete exhaustive surveys (such as RFGS) this is understandable.\footnote{There are many
challenges associated with assembling an exhaustive list.  A small number of books are published but
never advertised in industry publications such as \emph{Publishers' Circular}. In other cases,
novels may be advertised but never published, or published under a different title. Bibliographic
work is complicated further by the fact that in a (very) small number of cases, no copies of a novel
survive.} The absence of information about novels published after 1836 is regrettable because this
period witnesses the rise of mass literacy and sees the publishing industry adopt practices and
organizational structures characteristic of the modern text industry
\parencite[328-329]{raven2007business}. What little information we have about the population of
literary works published after 1836 relies on inferences drawn from the heterogeneous population of
published books (novels and non-novels, new and reissued)
\parencites{weedon2003victorian,eliot1997patterns}.  Even here, however, the information is not
detailed enough to allow us to estimate the number of novels (new or reprinted), published during
any year or decade.

In this paper we estimate rates of novelistic production for each year between 1789 and
1919 from five existing data sources using a probabilistic model. In addition to annual publication
counts, the data permit us to estimate the proportion of new titles associated with men and women
authors.  Although we do not directly observe the number of new novels published in any year after
1836---or new novels by author gender after 1829---we infer credible intervals through the use of a
model of several correlated time series. Our results make visible, for the first time, a period of
particularly intense growth between \periodOfGrowthStart{} and \periodOfGrowthEnd{}.

\subsection*{Background}

There are bibliographies and related resources that purport to provide information about new novels
published during specific periods of the 19th century. Most are unusable.  Typical are
bibliographies of a period or novel subgenre which for one reason or another are not exhaustive.
\textcite{block1961english} is one example. Although it advertises itself as a bibliography of
English novels published between 1740 and 1850, it is not clear what novels are included and what
novels are missing. Worse, it includes books which are not novels by any prevailing definition
\autocite[2]{garside2000general}.  There are, however, a small number of works which are exhaustive
for a period or genre and do provide information usable by those interested in an inclusive history
of the novel and of novel writing.  \textcite{bassett2008production}, for example, enumerates
all three-volume editions appearing between 1863 and 1897. RFGS, mentioned earlier, enumerates all
novels published between 1770 and 1836. RFGS also helpfully makes clear how they go about the
essential task of distinguishing novels from non-novels \parencite{garside2000general}.

For those interested in an inventory of new novels published during the 19th century, the most useful
information comes from historians of publishing. (With notable exceptions---including
\textcites{escarpit1958sociologie,moretti1982anima,moretti1998atlas,moretti2000slaughterhouse}---literary
historians working after 1950 have not pursued an inclusive history of the novel, one which would
include all novels and novelists.) Working with a machine-readable version of the \emph{Nineteenth Century Short Title Catalog}
(NTSC), \textcite{eliot1997patterns} creates a time series which provides information about the number of books
published in London, Oxford, Cambridge, Edinburgh, and Dublin each year between 1801 and
1870.\footnote{Working with data from \textcite{eliot1994patterns}, \textcite{weedon2003victorian}
combines the work of Eliot with other sources to offer a succinct description of publishing between
1836 and 1919 \parencite[46-51]{weedon2003victorian}.} Until an integrated history of the English
novel and the book trade is written, this series will be invaluable. It helps us in two specific ways.
First, it provides a crude upper bound on the number of new novels published each year as the number of
new novels will always be less than the number of books (novels and
non-novels) appearing in a given year. Second, because
the rate of book production and the rate of new novel production are correlated, the time series
gives us considerable insight into how the rate of new novel production likely changed from year to
year.

The two most important resources used to estimate the rate of novelistic production are RFGS and a
series derived from the \emph{Nineteenth-Century Short Title Catalog} (NSTC). Three other resources
used in the model---which tend to cover shorter periods---are introduced in the next section.

\subsection*{Method}

We estimate annual rates of novelistic production from five data sources using a probabilistic
model.  The model assumes that changes in the pace of novelistic production are well described by
exponential growth with transitory deviations. Using the model and available data we infer the pace
of growth and the character of deviations. Taken together these inferences permit us to estimate the
number of novels published each year between 1789 and 1919. In this section we first describe the
resources used and then elaborate the model.

\paragraph{The English Novel, 1770-1836 (``RFGS'')} The most important source of information is
\emph{The English Novel}, an exhaustive survey of novels appearing between 1770 and 1836
\parencites{raven2000english, garside2000english, garside2006english}. In this paper we refer to the
two-volume printed bibliography, updates, and online database collectively as RFGS.\footnote{%
  To the best of our knowledge, \textcite{garside2006english} includes corrections and additions to
  \textcites{raven2000english, garside2006english} which have been published online from time to
  time (e.g., \textcite{garside2001english}).
}

RFGS anchors the analysis in this paper in several respects. What RFGS records, counts of new
novels---and, for 1800-1829, counts by author gender---is what we wish to infer for the entire
period (1789-1919). RFGS provides a principled, descriptive definition of the novel: printed works
referred to as novels by readers at the time. The usefulness and specificity of this definition is
amplified by the fact that RFGS provides examples of works which meet the definition (the
bibliography itself) as well as works which do not meet the definition.  RFGS includes detailed
records for each title listed in the bibliography. For years 1800--1829, each record includes an
indication of the gender of the author. RFGS code author gender as (``Male'',``Female'',
``Unknown'').  If the title indicates author gender but not author name, the title is associated
with the indicated author gender.  For example, although the novel \emph{The Castle of Probation}
(1802) does not have a named author, it is associated with a ``Male'' author in RFGS because the
novel's full title includes the words ``By a Clergyman''.\footnote{``THE CASTLE OF PROBATION, OR,
PRECEPTIVE ROMANCES; CHIEFLY TAKEN FROM LIFE. BY A CLERGYMAN. IN TWO VOLUMES.'' (RFGS record no.
1802A002).} As a practical matter, we see RFGS as providing two distinct time series: first, counts
of new novels published between 1770 and 1836; and, second, counts of new novels by author gender
between 1800 and 1829. We further limit our attention to records associated with 1789 and later
years in order to allay concerns about the definitional strategy used. As the 18th century
progresses, characteristics associated with works labeled ``novels'' tend to stabilize. Works
published after 1789 which were referred to as novels are very likely to share morphology with works
labeled novels published during later decades. This is less often the case for novels published
earlier in the 18th century.

To address the concern that the definition used by RFGS may be too restrictive, that it may tend to exclude
literary works which were not called novels but which are, in all other respects, treated by readers
at the time as if they were novels, it is worth noting that different definitions of the novel tend
to agree on particulars in more than 85\% of cases. Moreover, disagreement is localized.  Most
disputed cases involve novel-like (didactic) juvenile fiction and novel-like religious fiction (Troy
Bassett, personal communication, Nov. 9, 2015). It should, therefore, be straightforward for other
researchers to adjust the estimates reported here or to modify the model source code accompanying
this paper to accommodate different assumptions about what works count as novels.

\textbf{Nineteenth-Century Short Title Catalog (London, Oxford, Cambridge, Edinburgh, or Dublin),
1801-1870 (``LOCED'').}
\textcite{eliot1997patterns} extracts yearly totals of entries (novels and non-novels) listed in
the \emph{Nineteenth-Century Short Title Catalog} (NSTC) associated with
one of the following places of publication: London, Oxford, Cambridge, Edinburgh, or
Dublin. We refer to this time series using Eliot's abbreviation, ``LOCED''. Because RFGS provide an exhaustive
survey of new novels between 1801 and 1836, we know what percentage of LOCED titles are new novels
for 36 years. During these years there is an opportunity to observe how the two time
series covary.

Our LOCED series differs from Eliot's in one important respect.  The original LOCED series has an
unusual feature: undated material is assigned to the nearest half-decade (to a year ending with a
``0'' or a ``5'') \parencite[86]{eliot1997patterns}. To deal with this idiosyncrasy, we ignore
entirely publication counts from the original series which are associated with years ending in ``0''
or ``5''. Although ignoring counts in these years might appear to bias the counts associated with
other years downward (as many works, were their publication years known, ``belong'' in adjacent
years), we have a different view. The original LOCED series mixes two time series, a series
recording dated material and a series recording undated material. (New novels, for example, are
virtually certain to report publication years on their title pages.) By stripping out counts for
years ending with ``0'' or ``5'', we ignore the time series related to undated
publications.\footnote{Of course, ignoring counts in years ending with ``0'' or ``5'' means
discarding potentially useful information about counts of dated publications.  Separating counts of
dated material from undated material in these years would be valuable.}

\paragraph{Publishers' Circular, 1843-1919 (``PC'')} The third time series we use records yearly
totals of new titles derived from \emph{Publishers' Circular}, 1843--1919
\parencite{eliot1994patterns} (``PC''). Issues of \emph{Publishers' Circular} appeared biweekly and
listed new books published.  The PC time series overlaps with LOCED for 28 years (1843--1870),
permitting observation of how these two series covary. As one would expect given the similarity in
what is being recorded in the two series, the PC series and the LOCED series are highly correlated
($r = 0.72$). Together they give us a guide to year-to-year variation in the rate of book
publication over 119 years (1801--1919).

At this point the inference strategy may be growing clearer. We aim to gather several partially overlapping
time series which are correlated in order to ``triangulate'' from observed rates to unobserved rates.

\paragraph{\emph{The Athenaeum} Reviews of Novels, 1860, 1865, \ldots, 1900} The fourth and fifth
resources are used primarily to improve the estimates of the number of new novels published after
1850. Improving our estimates for this period is important because uncertainty grows as we move
further away from the bibliographic terra firma of the early 19th century. The fourth resource
appears in \textcite{casey1996edging}. Casey provides counts for the number of novels reviewed in
\emph{The Athenaeum} during nine years: 1860, 1865, 1870, 1875, 1880, 1885, 1890, 1895, and 1900.
(\emph{The Athenaeum} was a London literary magazine published from 1828 to 1921.) Casey also breaks
down the number of novels reviewed during the nine years by author gender. We make the assumption
that every title counted as a novel in this time series meets the definition of a novel used by
RFGS.

Counts are taken from Chart 2 in \textcite{casey1996edging}. In Casey's series, titles with multiple
authors contribute an author fraction to the relevant count. As the model used here is designed to
model count data, all non-integer values in Chart 2 are rounded down. As novels with multiple
authors are exceedingly rare during the period, we feel that ignoring authors other than the first
will not meaningfully change any results presented in our analysis.

\emph{The Athenaeum} does not review all novels published, so these counts are significantly lower
than the total number of new novels published. If we knew the percentage of new novels reviewed by
the magazine, we could derive the number of new novels published during these nine years.  We infer
the percentage of novels reviewed by modeling the overlapping time series. This strategy is the same
as the one used to infer the percentage of total books published which are novels. In our model, we
assume that the percentage of novels reviewed, whatever it turns out to be, is fixed during the
period 1860--1900. Supporting this assumption is the observation that novel reviews in \emph{The
Athenaeum} increased markedly between 1860 and 1900, suggesting that the periodical enjoyed
flexibility in the number of titles it reviewed.

\paragraph{Elicited Distributions of New Novel Publications in 1886, 1891, and 1894} The fifth
resource, like the fourth, is used to reduce the considerable uncertainty about the number of new
novels published in the second half of the 19th century. The fifth resource is a series of three
\emph{distributions} over rates of new novels publication in the years 1886, 1891, and 1894. These
distributions are elicited from a domain expert, Troy Bassett, editor of \emph{At the Circulating
Library: A Database of Victorian Fiction, 1837-1901} (``ATCL'').\footnote{%
  These years were chosen because a preliminary model made implausible predictions for
  these years. The predictions were implausible in that they were near or lower than a lower bound on the
  number of novels published in the relevant years. Lower bounds were available for these years
  because the ATCL database already contains records for many thousands of novels published in the
  19th century.
}
We follow the elicitation procedure described in \textcite{garthwaite2005statistical}. For each
year, we asked Bassett to report quartiles of the distribution reflecting his beliefs about the
total number of new novels published that year. As editor of ATCL, a database which contains entries
for over 15,000 novels published between 1837 and 1901, Bassett is in a position to make accurate
estimates of intervals which are likely to contain the total number of new novels published in any
year during the Victorian period.  Eliciting quartiles of a distribution which describe the likely
number of new novels published in a year is roughly equivalent to asking for an interval which
contains the true number with probability 0.5.  After eliciting quartiles of the distributions for
the three years, we find familiar probability distributions which have quartiles as close as
possible to those elicited.  The three distributions identified in this way are the distributions
used in the model. For example, the quartiles elicited for the year 1886 are 394, 482, and 613. A
normal distribution with mean 494 and standard deviation
163 has approximately the same quartiles: 384, 494, 604.\footnote{%
  The distributions were elicited in a phone conversation between Allen Riddell and Troy Bassett on
  November 9th, 2015.  The quartiles reported in the paper are discounted from the original
  quartiles (450, 550, 700). Discounting is required because ATCL uses a more inclusive definition
  of the novel than RFGS. (For example, RFGS exclude some religious and didactic fiction that ATCL
  includes.) Bassett reports that between 10\% and 15\% of the novels included in ATCL would not be
  counted as novels according to RFGS. For this reason we discount the reported quartiles by 12.5\%
  (the midpoint between 10\% and 15\%). The matching of ideal distributions to the elicited
  distributions (implied by the quartiles) involves one additional step because we model the rate of
  new novel publication on the log scale. We use Gamma distributions which have quartiles as close
  as possible to the elicited distributions (now on the log scale). For example, the final
  representation of the distribution with quartiles 394, 482, and 613 is (on the log scale) a Gamma
  distribution with shape and rate parameters of 278 and 46.
}

\subsubsection*{A Model of Novelistic Production}

In this section we review the most important assumptions we make in our
model---exponential growth with transitory deviations---and then describe in detail how the five
time series mentioned earlier appear in the full model. To simplify the
presentation, we initially describe the model without considering author gender. The minor
adjustments required to model author gender are presented at the end of this section.

Seen from a distance, it is obvious that the rate at which new novels appear grows exponentially.
We can appreciate this by looking at the rate at which books
(novels and non-novels) appear \parencites{eliot1997patterns}{weedon2003victorian}. Additional
evidence, if any is needed, is available from \textcite{eliot1998patterns} which shows nonlinear
growth in the number of titles labeled as ``Literature'' in the NSTC
\parencite[85]{eliot1998patterns}. The standard approach to modeling this sort of trend is a
log-linear model. Taking log publication rates as our estimands, we can describe the trend using
a linearly increasing rate of publication. In a log-linear model, the log rate of new novel
publication in year $t$ is described by a two-parameter expression, $\alpha + \beta t$,
where $\beta$ is interpreted as an annual growth rate. (For example, if in year 1800 the
annual rate of publication is 100 new novels and the rate grows continuously at a rate of 3\%,
$\beta$ would be 0.03 and in the year 1900 the annual rate of publication would be roughly 2,000 new
novels.) In our model of the log rates of new novel publication, a linear trend appears as the mean
function of a Gaussian Process.

Both the linear trend and transitory deviations are modeled by a Gaussian Process.
A Gaussian process allows us model growth of new novel publication
using a simple exponential trend while also permitting us to account for transitory deviations from this trend due to disruptions in the
book trade (e.g., economic depressions, wars, cholera outbreaks, and so forth).
The time series derived from the NSTC (LOCED) and \emph{Publishers' Circular} (PC) make clear that novel publishing
experienced several disruptions between 1789 and 1919. Time series of the number of books published
suggest the influence of events including wars, market panics, and epidemics. That the disruptions
are transitory is also clear. The text industry always returns to growth. Because Gaussian Processes
can model both an underlying trend and transitory deviations, they are a familiar choice
in settings similar to this one. (As Gaussian Processes are covered in detail elsewhere---for example, in
\textcite{rasmussen2005gaussian} and \textcite{bishop2007pattern}---we do not describe them in any
detail here.)
The backbone of our model is therefore a Gaussian Process of the log rate of new novel publication between 1800 and 1919. In
symbols, the log rate of new novel appearance for year $t = 1, \ldots, 120$ is given by

\begin{align}
\lambda_t &= \alpha + \beta t + \epsilon(t),\\
\epsilon(\vec t) &\sim \text{GP}(0, K) \\
k(t, t^\prime) &= \sigma_\lambda^2 \exp\left(-\frac{|t - t^\prime|^2}{l_\lambda^2} \right)
\end{align}
where the year $t = 1$ is associated with 1800, $t = 2$ with 1801, and so on. $\text{GP}(0, K)$ is a
zero-mean Gaussian Process with $120 \times 120$ covariance matrix $K$; and the element $(t, t^\prime)$
of $K$ is given by $k(t, t^\prime)$.

Two examples may help make the covariance matrix $K$ more intelligible. $K_{2,3}$ is the covariance between the observation $\lambda_2$, the log rate for 1801, and $\lambda_3$, the log rate for 1802.
Its value is $\sigma_\lambda^2 \exp\left(-\frac{|2 - 3|^2}{l_\lambda^2} \right)$.
$K_{2,120}$ is the covariance between the rate for 1801 and 1919.
Unless $l_\lambda$ is extremely small, $K_{2,120}$ will be near zero because it contains the term $\exp\left(-\frac{|2 - 120|^2}{l_\lambda^2} \right)$.
($\exp(-c)$ will be near zero whenever $c$ is a large number.)
A near-zero covariance makes sense here because we do not anticipate an observation of the 1801 rate telling us anything about 1919 rates.

To capture the belief that deviations from the trend will
tend to persist for a bounded number of years, we use an informative prior distribution on the
characteristic length-scale $l_\lambda$.  This distribution places 90\% probability on values
between 1 and 10, expressing the prior belief that deviations will tend to persist for between 1 and
10 years.  Such a prior distribution is consistent with the belief that, say, a market panic might affect
the rate of novel publication in the short term but would likely cease to influence publication
rates in years which are more than ten years distant from the event.
Here, as elsewhere, we draw on domain expertise to justify our modeling choices.
Different choices will lead to different results.
(Different models—say, linear or quadratic rather than log-linear—will lead to radically different results.)
Readers who prefer different assumptions are invited to edit the code which accompanies this article and develop models which reflect their beliefs.

The observed annual counts of new novels from RFGS (1800-1836) (the first time series) are connected
to the latent log rates $\lambda_{1:37}$ via a negative binomial sampling distribution. This sampling
model allows us to connect the smoothly varying rates to observed counts of new novels.  Separating
the latent rate from the observed counts in the model is particularly important before 1840 because
there is considerable year-to-year variation in the observed counts of new novels which are due to
the arbitrary assignment of novel publications into discrete years.\footnote{%
  One way of appreciating the importance of modeling new novel publication with a continuous rate
  parameter is to imagine a situation where the aleatory variation in new novel counts is
  considerably greater. Imagine modeling new novel publication via weekly counts. In such a setting
  observing that zero new novels appeared in a given week would not be particularly meaningful. It
  would certainly not imply that there was zero activity associated with novel publishing during
  that week.
}
In symbols, the sampling model is given for year $t = 1, \ldots, 37$ by
\begin{align}
y_t \sim \text{NegativeBinomial}_2(\exp(\lambda_t), \phi_y)
\end{align}
where $\text{NegativeBinomial}_2$ is parameterized by
a location parameter and a parameter controlling dispersion. (If $Y$ is distributed according to a
$\text{NegativeBinomial}_2(\mu, \phi)$ distribution then $\E(Y) = \mu$ and $\Var(Y) = \mu +
\frac{\mu^2}{\phi}$.) We use a two-parameter negative binomial sampling model here rather
than a simpler, single-parameter Poisson model. The former's ability to model additional variation
is important given the uncertainty about the latent process being modeled.

To incorporate the counts of \emph{Publishers' Circular} (PC) titles (the second time series), we
introduce an additional Gaussian Process to model, for each year, the proportion of PC titles which
are new novels. Background knowledge and \textcite{eliot1998patterns} lead us to believe that the
proportion will be certainly less than 50\% and that it will increase modestly over the period.  As
we did for the rates of new novel appearance, we transform the proportions into units which are
conveniently modeled using a linear trend. In this case, we express the proportions on the log odds
scale, denoting the log odds as $\nu_t$ for year $t$. (The log odds is the logarithm of the odds,
$\log(\frac{p}{1-p})$, where $p$ is a proportion between 0 and 1.)  In contrast to our thinking
about year-to-year variation in rates of new novel publication, we anticipate that the proportion of
PC titles which are new novels will change comparatively slowly.  Whereas an economic crisis or
other kind of ``shock'' might affect the rate of new novel publication over a period of several
years, it would likely not affect the proportion of books which are novels. In other words, we
anticipate that factors influencing the economics of publishing novels as opposed to non-novels does
not change as rapidly as factors influencing the rate of book publishing in general. To capture this
belief, the characteristic length-scale for this second Gaussian Process is modeled with a prior
distribution placing 90\% probability on values between 8 and 36, expressing the belief that
deviations from trend will tend to persist for between 8 and 36 years. In symbols, the proportions
are modeled for year $t = 1, \ldots, 120$ on the log odds scale as follows:

\begin{align}
\nu_t &= \alpha_\nu + \beta_\nu t + \epsilon_\nu(t)\\
\epsilon_\nu(\vec t) &\sim \text{GP}(0, K_\nu) \\
k_\nu(t, t^\prime) &= \sigma_\nu^2 \exp\left(-\frac{|t - t^\prime|^2}{l_\nu^2} \right)
\end{align}

As with the yearly novel publication counts, observations of PC title counts (1843-1919) are
connected to latent rates via a negative binomial sampling distribution. The latent rate of PC title
appearance in year $t$, the mean of the sampling distribution, is $\exp(\lambda_t) /
\logit^{-1}(\nu_t)$, where $\logit^{-1}$, the inverse logistic function, is the inverse of the
transformation of a proportion into log odds.  For example, if the proportion of PC titles which are
novels is 12\% and the rate of new novel appearance is 300 then the observed PC title count will be
modeled with a negative binomial distribution with mean 2,500.

The yearly \emph{Nineteenth-Century Short Title Catalog} (LOCED) publication counts (the third time
series) record similar information as the PC title counts series. They both record total publications (novels and non-novels).
They differ primarily in the years they cover.
The PC counts tend to be lower because PC tends to only report editions for sale in London.
Because these series are very similar, we model the
LOCED rate in terms of the PC rate. We assume that the LOCED rate is a fixed multiple of the PC
rate. The rate at which titles are recorded in LOCED is incorporated into the model by assuming that
the rate is the same as the PC rate, multiplied by a constant factor, $\pi_\nu$. Because LOCED
counts are always greater than PC counts, this factor will be greater than one.\footnote{
  Counts derived from the NSTC and PC supply essential quantitative information about the development of text industry in the British Isles and Ireland.
  In particular, these time series provide information about the year-to-year variation in the number of editions produced by the text industry.
  These sources have been used in previous research and are certain to be used in the future.
  While a precise understanding of their relationship is a topic for another paper, we can offer some preliminary observations.

  We know that for any given year the PC series always reports fewer editions than LOCED.
  The reason for this is, we suspect, that PC tends to only report titles for sale in London.
  LOCED, by contrast, contains records for all editions which ended up in libraries.
  Since there was a legal deposit requirement and LOCED includes records from the legal deposit libraries, LOCED covers a broader range of editions.
  LOCED gives us a sense of all editions published in the British Isles and Ireland, not just those published or distributed in London.
  For example, technical works published by university presses in Oxford, Cambridge, and Edinburgh which were not distributed in London would likely appear in LOCED.
  These editions would tend not to appear in PC.

  In our model we assume that, for every year, the number of editions in LOCED is a fixed multiple of the number of editions in PC.
  We make this assumption because it simplifies the model and because we think it is a reasonable assumption.
  It is a reasonable assumption if one believes that the rate of growth of publishing outside of London grew at the same rate as publishing in London.
  The reasoning behind such a belief should be familiar at this point. %
  Technological changes in the text industry such as cheaper paper and cheaper printing shaped publishing everywhere, not just in London.
  The same holds for relevant institutional changes, such as lower costs of capital associated with maturing financial institutions.
  So the fixed multiple assumption rests on the belief that the PC series captures the number of titles for sale in London and LOCED captures the number of titles published in London as well as in publishing centers outside of London.
  If the rate of publishing grew at the same pace throughout the British Isles and Ireland, the ratio of LOCED titles to PC titles should be approximately constant.
} As before, a
negative binomial sampling distribution connects this yearly rate to the observed LOCED counts
(1801-1870). For reasons discussed earlier, LOCED counts from years which end in a '0' or '5' are
ignored.

Counts of new novels reviewed in \emph{The Athenaeum} (the fourth time series) are incorporated into
the model using a similar strategy to the one just described for LOCED title counts. The rate at
which novels are reviewed is assumed to be equal to the rate of new novel publication multiplied by
a constant factor, $\pi_a$. The use of a constant factor reflects the assumption that the proportion
of new novels reviewed in \emph{The Athenaeum} was roughly the same during each of the nine years.
As noted earlier, that \emph{The Athenaeum}'s reviewing expands considerably during the
period (from 137 in 1860 to 473 in 1900) lends this assumption superficial plausibility. As we
know in advance that \emph{The Athenaeum} does not review all new novels, an informative Gamma
prior distribution placing 90\% probability on a value between 30\% and 70\% is used.  As with the
other count-based time series, a negative binomial sampling model is used to model the relationship
between latent rates and observed counts.

We connect the three distributions elicited from Bassett (the fifth data source) directly to new
novel publication log rates for the relevant years ($\lambda_{87}$, $\lambda_{92}$, and $\lambda_{94}$).
This makes incorporating the distributions into the model straightforward: the three elicited
distributions are used as prior distributions on the rate of new novel appearance during 1886, 1891,
and 1894. Although a meticulous approach would associate the three distributions with the unobserved
\emph{counts} of new novel publications---this is, after all, what Bassett was asked about---such an
approach would add considerably complexity to the model by requiring us to model latent discrete
variables (the unobserved counts). Assuming that the Bassett estimates concern continuous latent
rates rather than discrete counts has the consequence of modestly understating the variance of the
elicited distributions. Given that the elicited distributions indicate a generous degree of
uncertainty we think this is a reasonable price to pay for a simpler model.

\paragraph{Modeling author gender} The essential structure of the model has been introduced. The
full model differs slightly from the version presented. In addition to estimating the number of new
novels published each year, the full model also estimates the number of novels published by author
gender. This is accomplished by adding, for each year, two parameters to the model. The first
parameter, $\rho_t$, records the proportion of new novels associated with an author of unknown
gender. The second parameter, $\sigma_t$, records the proportion of known-author-gender new novels
associated with men authors (a proportion of a proportion). With these two parameters it is possible
to calculate the proportion of new titles given each of the three author gender annotations.  For
example, new novels associated with women authors in year $t$ is given by $(1 - \rho_t) (1 -
\sigma_t)$. Each sequence, $\rho_{1:120}$ and $\sigma_{1:120}$, is modeled on the log odds scale
using Gaussian Processes with a linear trend. Prior distributions for the characteristic
length-scale parameters are the same as the prior distribution used for the length-scale parameter
for the Gaussian Process model of $\nu_{1:120}$ (the proportion of PC titles which are new novels).
Observed counts of new titles by author gender---available in \emph{The Athenaeum} series and, for
1800 to 1829, in RFGS---are modeled with negative binomial sampling distributions.

\paragraph{New novels by author gender, 1789-1799}
We estimate the number of new novels by author gender separately for the 11 years between 1789 and
1799. Because the number of new novels published during this period appears in RFGS, we need only
estimate, for each year, the proportion of novels associated with men, women, and unknown gender
authors. We accomplish this by collecting and manually annotating a random sample of 110 titles from
RFGS (ten titles for each year). For each year we calculate a posterior distribution over
proportions using a multinomial sampling model and an informative Dirichlet prior distribution
loosely centered on observed proportions in 1800.

For the full model covering the period between 1800 and 1919, we estimate model parameters using Markov Chain Monte Carlo \parencite{carpenter2017stan}. (For a general introduction to Monte Carlo methods in Bayesian statistics see \textcite{liu2002monte}.)  All parameters whose
prior distributions are not discussed are given reasonable, weakly informative prior distributions.

\subsection*{New Novel Publications, 1789--1919}

\begin{figure}
  \includegraphics{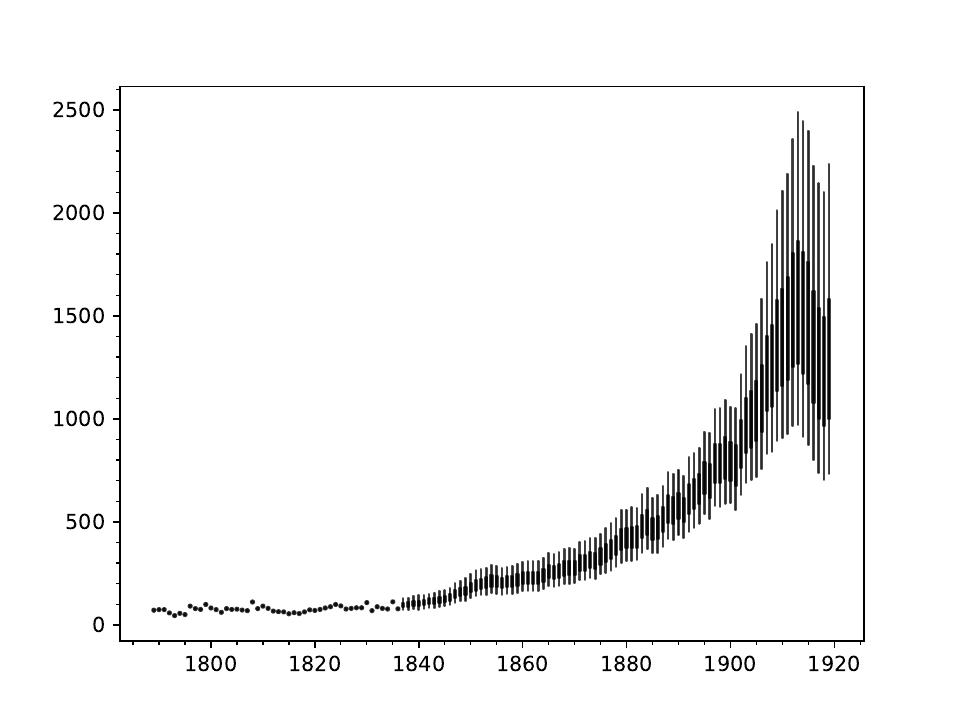}
  \caption{\textbf{New novels, 1789--1919.}
  Figure shows new novels published in the British Isles and Ireland between 1789 and 1919.\label{fig:new-novels}}
\end{figure}

\begin{figure}
  \includegraphics{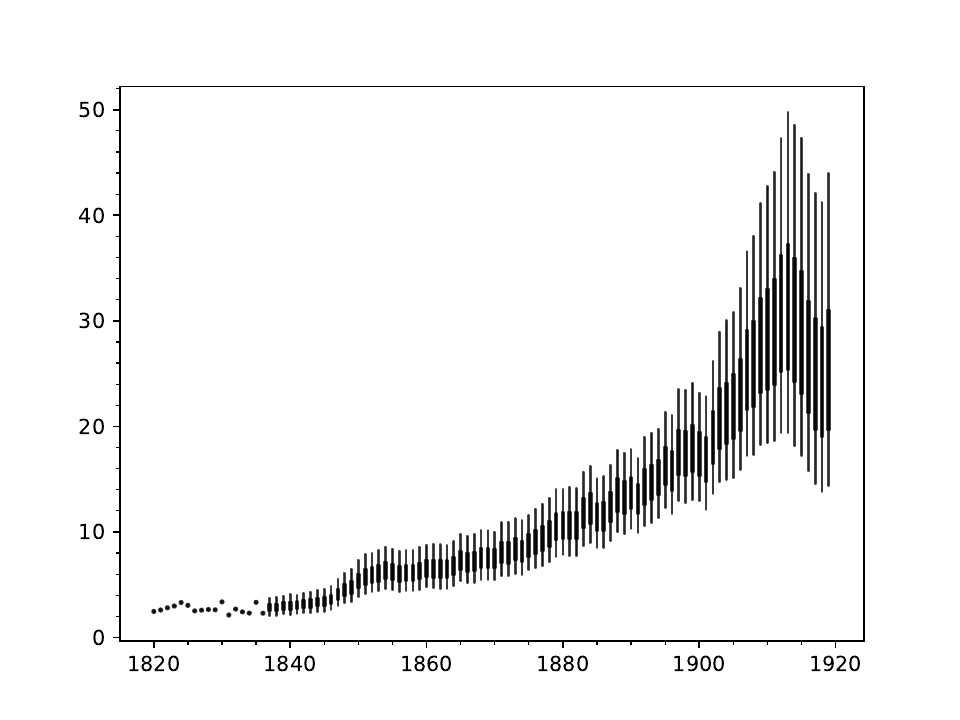}
  \caption{\textbf{New novels per million persons, 1820--1919.}
  Figure shows new novels published in the British Isles and Ireland, per million persons, between 1820 and 1919.
  Population figures are from \textcite{maddison2009statistics}. Population of the British Isles and Ireland is
  calculated by adding UK and Ireland populations. This series begins with 1820 because yearly population estimates are not available for earlier years. \label{fig:new-novels-per-person}}
\end{figure}

The model provides estimates of the rate of novel publications for each year between 1789 and 1919.
Figure~\ref{fig:new-novels} visualizes these rates. (Figure~\ref{fig:new-novels-per-person} shows these rates normalized by population.)
Each interval in Figure~\ref{fig:new-novels} shows the posterior credible interval for the rate of new novel publication, $\exp(\lambda_t)$, for a specific year $t$.
Points represent the number of new novels published during 1789–1836—a period for which we have exhaustive bibliographies.
In aggregate between
\NewNovelsSeventeenHundredEightyNineToNineteenHundredNineteenFifthPercentile{} and
\NewNovelsSeventeenHundredEightyNineToNineteenHundredNineteenNinetyFifthPercentile{} new novels
likely appeared between the years 1789 and 1919. (All intervals mentioned are 90\% credible
intervals.) A summary by decade appears in Table~\ref{tbl:novels-by-decade}.  For comparison, the
number of these titles which are still in print today is shown, by author gender and decade of
publication, in Table~\ref{tbl:novels-reprint-canon-by-decade}. This ``reprint canon'' (borrowing
the label from \textcite{bassett2017median}) serves as an approximation of the body of works
currently taught in universities. The reprint canon very likely represents less than one percent of
novels published during the period. It is possible that it represents as little as
one half of one percent of published titles.\footnote{
  The period between 1800 and 1899 is often the focus of discussion. Between
  \NewNovelsEighteenHundredToEighteenHundredNinetyNineFifthPercentile{} and
  \NewNovelsEighteenHundredToEighteenHundredNinetyNineNinetyFifthPercentile{}
  appeared between 1800 and 1899. Totals for other intervals may be calculated
  using annual publication rates shown in Table \ref{tbl:novels-by-year}.
  Table~\ref{tbl:novels-reprint-canon-by-year} shows reprint canon titles by
  author gender and year.}

\begin{figure}
  \includegraphics{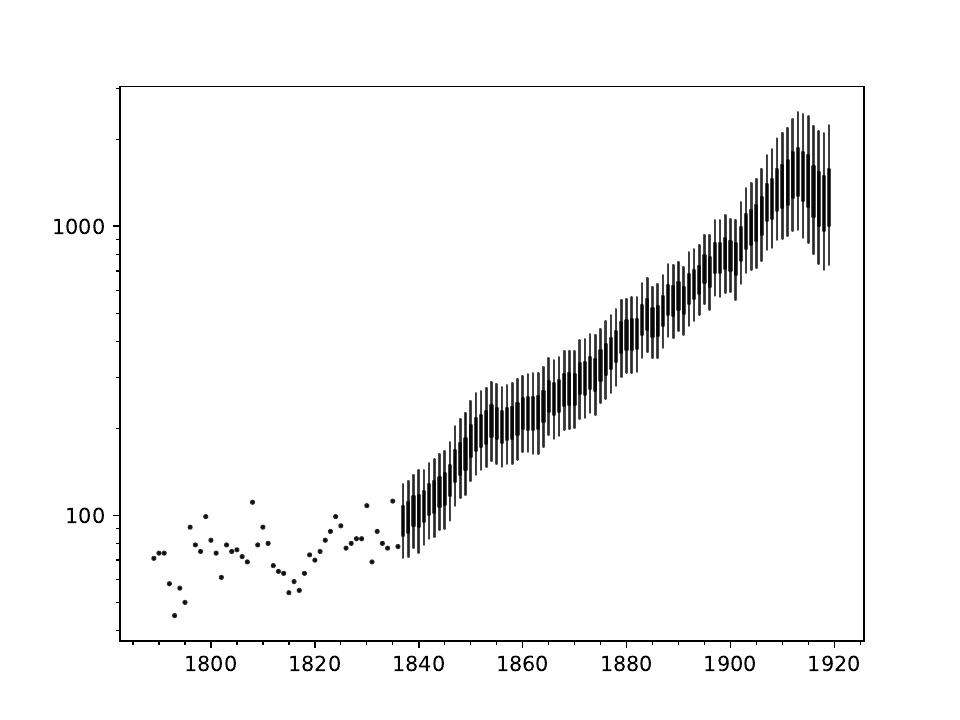}
  \caption{\textbf{New novels, 1789--1919 ($\log_{10}$ scale).}
  Figure shows new novels published in the British Isles and Ireland between 1789 and 1919 using a
  $\log_{10}$ scale on the vertical axis.\label{fig:new-novels-log10}}
\end{figure}

One remarkable development which is visible by inspection is the rapid growth in new
novel publication between \periodOfGrowthStart{} and \periodOfGrowthEnd{}.
Figure~\ref{fig:new-novels-log10} shows a plot the number of new novels published on a logarithmic scale.
In this figure three regimes of growth in the 19th century are clearly visible. Before
\periodOfGrowthStart{} there appears to be modest growth in the number of titles published each
year.  Between \periodOfGrowthStart{} and \periodOfGrowthEnd{} there is rapid growth in the number
of new titles produced. Average annual growth
during this period is 5\%.  Between \periodOfGrowthEnd{} and 1900 there is likely steady, but
markedly slower growth. The average annual growth during this period is likely 2\%. The rapid growth
during the \periodOfGrowthStart{}--\periodOfGrowthEnd{} period merits further
investigation. How did it come about and how was it sustained? What consequences did it have for the
network of actors involved in the literary market?  The rate of novel publication likely
doubled in the space of a 15 year period, requiring a rapid expansion in a range of processes of
interest to literary historians and historians of publishing. For example, this growth suggests a
doubling of the labor of compositors, a doubling of paper used, and a doubling of the
rate at which manuscripts were developed for publication. How was this rate of growth sustained? Did
one particular novel subgenre, group of intermediaries, or cohort of novelists benefit from this
expansion?  The rapid pace of growth seems likely to have left traces in a variety of places, not
least in the lives of writers and in the morphology of literary texts.

\begin{figure}
  \includegraphics{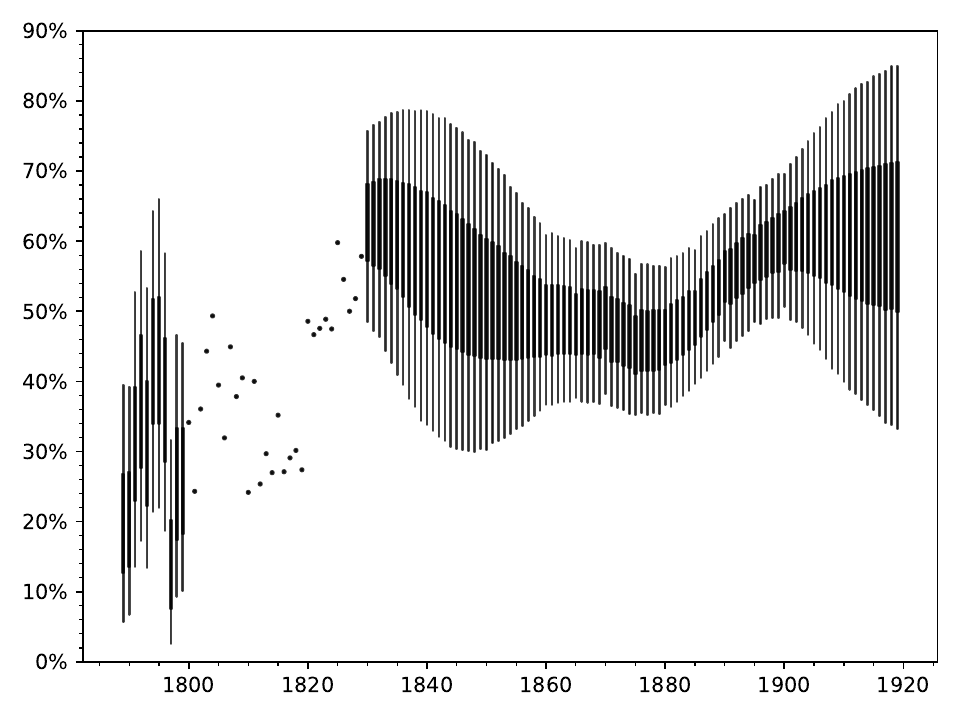}
  \caption{\textbf{Men authors' share of new novels, 1789--1919.} Percentage of new novels with known
  author gender which are novels by men authors. For years in 1800-1829 exhaustive gender
  annotations are known. For all other years model estimates are shown, with thick vertical bars
  indicating 50 percent credible intervals and thin vertical bars indicating 90 percent credible
  intervals.\label{fig:new-novels-gender-by-year-men-of-known}}

\end{figure}

Estimates of men authors' share of new novel publication by year is shown in Figure~\ref{fig:new-novels-gender-by-year-men-of-known}.
The estimates are consistent with the widely held
belief that there was a demographic shift in the occupation of novel writing during the 19th century
\parencite[5-11]{tuchman1989edging}. At the beginning of the 19th century a majority of novels with
known author gender were associated with women novelists. By the end of the 19th century this
percentage had likely declined to roughly 40\%.\footnote{%
  Our estimates concern the characteristics of the population of new novel titles, not novelists.
  If one assumes that novelist gender is uncorrelated with the number of
  novels they publish, then the share of novelists associated with each gender should be roughly the
  same as the share of novels associated with each gender.  Estimating the demographic
  characteristics of the population of professional novelists should be addressed in subsequent
  research. This research may need to, for example, avoid double-counting novelists who used
  different---or even collective---pseudonyms.
} Within the expected secular decline in the proportion of novels associated with women authors
there is some evidence of a cyclical trend: the proportion of titles associated with men authors declines
during the 1860s and 1870s before recovering again.\footnote{%
  \textcite{moretti2005graphs} suggests a connection between author gender and
  literary cycles during the 19th century.  Moretti, however, does not appear
  to credit the possibility of a long-term secular decline in the proportion of
  novels written by women authors \autocite[27]{moretti2005graphs}.
}

The estimates also permit us to say that it is virtually certain that novels by men authors and
novels first published in the 1860s are overrepresented among titles which are still in print today.
That is, the
proportion of novels associated with men authors in the reprint canon does not reflect the
proportion of novels written by men during the period. It is very likely that between 40\%
and 58\% of novels written between 1789 and 1919 were associated with men authors (Table
\ref{tbl:novels-by-decade}). In the reprint canon, however, 71\% of novels from this period are
associated with men authors (Table \ref{tbl:novels-reprint-canon-by-decade}). The distribution of
reprint canon titles by year of first publication is also not aligned with the distribution of
titles published during the period.  Titles published in the 1860s, in particular, appear to be
overrepresented in the reprint canon. Titles published in the 1900s appear to be underrepresented.
Although it is possible that the reprint canon does not reflect literary works used in research and
taught in university classrooms, the reprint canon does reflect the population of 19th century
novels which continue to be sold and read.

\begin{table}
  \begin{tabular}{lllll}
\toprule
{} &            Men-authored &          Women-authored &              Unknown &              N \\
\midrule
1790-1799 &        171-256 (24-36\%) &        316-411 (45-58\%) &      91-161 (12-22\%) &            701 \\
1800-1809 &               298 (30\%) &               366 (40\%) &            114 (12\%) &            778 \\
1810-1819 &               197 (28\%) &               346 (42\%) &            126 (13\%) &            669 \\
1820-1829 &               426 (41\%) &               289 (30\%) &            114 (11\%) &            829 \\
1830-1839 &        355-611 (38-65\%) &        203-441 (21-46\%) &       88-185 (9-19\%) &      826-1,047 \\
1840-1849 &        355-846 (27-63\%) &        276-759 (21-56\%) &      126-331 (9-24\%) &    1,087-1,571 \\
1850-1859 &      570-1,195 (29-54\%) &      568-1,184 (29-53\%) &     221-531 (11-25\%) &    1,653-2,549 \\
1860-1869 &      791-1,331 (35-48\%) &      839-1,402 (37-51\%) &     254-495 (10-18\%) &    1,996-3,084 \\
1870-1879 &    1,080-1,795 (35-48\%) &    1,253-2,051 (41-55\%) &      247-457 (7-12\%) &    2,685-4,172 \\
1880-1889 &    1,743-2,783 (39-52\%) &    1,771-2,802 (40-52\%) &      270-516 (5-10\%) &    3,932-5,907 \\
1890-1899 &    2,892-4,520 (46-59\%) &    2,121-3,373 (33-45\%) &       399-715 (6-9\%) &    5,617-8,363 \\
1900-1909 &    3,925-8,111 (42-67\%) &    2,306-5,503 (24-47\%) &    577-1,649 (6-15\%) &   7,827-13,911 \\
1910-1919 &   3,922-13,001 (33-75\%) &    1,999-9,289 (16-56\%) &    552-3,432 (4-22\%) &   9,060-21,896 \\
All       &  18,330-33,490 (41-58\%) &  14,360-26,354 (31-47\%) &  3,701-8,033 (7-15\%) &  39,560-62,907 \\
\bottomrule
\end{tabular}

  \caption{\textbf{New novels published between 1790 and 1919.} Intervals show 90\% credible
  intervals. Percentages shown are calculated with respect to table rows. As the total number of new
  novels published between 1789 and 1836 is known, totals for the 1790s, 1800s, 1810s, and 1820s are
  reported. Similarly, exhaustive gender annotations are available for the years 1800-1836, so
  totals for the 1800s, 1810s, and 1820s are reported. For yearly estimates see
  Table~\ref{tbl:novels-by-year}.  \label{tbl:novels-by-decade}}

\end{table}

\begin{table}
  \begin{tabular}{llllr}
\toprule
{} & Men-authored & Women-authored & Unknown &    N \\
          &              &                &         &      \\
\midrule
1790-1799 &      5 (23\%) &       17 (77\%) &  0 (0\%) &   22 \\
1800-1809 &      4 (27\%) &       10 (67\%) &  1 (7\%) &   15 \\
1810-1819 &     10 (48\%) &       11 (52\%) &  0 (0\%) &   21 \\
1820-1829 &      6 (67\%) &        3 (33\%) &  0 (0\%) &    9 \\
1830-1839 &      5 (83\%) &        1 (17\%) &  0 (0\%) &    6 \\
1840-1849 &     16 (73\%) &        6 (27\%) &  0 (0\%) &   22 \\
1850-1859 &     15 (65\%) &        8 (35\%) &  0 (0\%) &   23 \\
1860-1869 &     23 (62\%) &       14 (38\%) &  0 (0\%) &   37 \\
1870-1879 &     30 (83\%) &        6 (17\%) &  0 (0\%) &   36 \\
1880-1889 &     32 (82\%) &        7 (18\%) &  0 (0\%) &   39 \\
1890-1899 &     39 (87\%) &        6 (13\%) &  0 (0\%) &   45 \\
1900-1909 &     34 (94\%) &         2 (6\%) &  0 (0\%) &   36 \\
1910-1919 &     12 (86\%) &        2 (14\%) &  0 (0\%) &   14 \\
All       &    231 (71\%) &       93 (29\%) &  1 (0\%) &  325 \\
\bottomrule
\end{tabular}

  \caption{\textbf{Novels published between 1780 and 1919 which are still in print.} The table shows counts of novels
  originally published between 1789 and 1919 are available from Broadview Press, Penguin, or
  Oxford in 2018. As no novels originally published in 1789 are still in print, the totals shown
  reflect the totals for 1789--1919. Sources: Broadview Press 2018 English Catalogue, Penguin
  Classics 2016 Catalog, Oxford World's Classics 2016
  Catalog.\label{tbl:novels-reprint-canon-by-decade}}

\end{table}

\subsection*{Limitations and Future Work}

The estimates presented here reduce uncertainty about the number of new novels published between
1789 and 1919.  The reduction is significant enough that a variety of existing narratives of
developments in the literary market and the text industry merit revisiting in light of the new
estimates. The account offered by \textcite{tuchman1989edging} of changes in the percentage of women
pursuing careers as novelists is one example. The census data Tuchman uses to gauge changes between
1861 and 1919 are, by her own admission, unreliable \autocite[58]{tuchman1989edging}.  Although the
estimates presented here concern the annual number of titles published by author gender and not the
number of working women novelists, the series presented here is more detailed and
more relevant to the quantities of interest to Tuchman than any series available in the 1980s.
Research on the social history of novel writing similarly merits
revisiting in light of these new estimates.\footnote{A reference point for this kind of research, in
addition to Tuchman, is \textcite{williams1965long}. Past studies have explored---often with
fragmentary or conspicuously partial or biased samples of writers---the social, educational, and
geographic background of writers \autocites[261-263]{williams1965long}[113-119]{tuchman1989edging}.}
New research here would potentially complement any investigation into periods of particularly
rapid growth in new novel publication (e.g., \periodOfGrowthStart{}--\periodOfGrowthEnd{}), as
the factors driving this expansion may be illuminated by studying the differences between cohorts of
writers before and after the expansion.

Although the estimates here give us greater confidence about the annual rates of new novel
publication, much work remains to be done. The estimated intervals are wide, especially after 1850.
Narrowing the intervals will require more precise information about novelistic production during the
late 19th and early 20th centuries. One simple, effective strategy for gathering such information
would involve conducting an exhaustive survey of novels published in a single year after 1850.
Because knowing the rate of production in a given year provides information about plausible rates
for neighboring years, accurate information about a single year would improve estimates of
neighboring years. Although collecting an exhaustive list of novels published in, say, 1865 would be
time-consuming, the work itself is straightforward: new novels need to be identified among all
entries in \emph{Publishers' Circular} for the chosen year.

The period of rapid growth between \periodOfGrowthStart{} and \periodOfGrowthEnd{}, if evidence for
its existence continues to accumulate, deserves further study. Did one or a small number of factors
drive this growth? Was the growth attributable to, for example, lower per-unit costs arising out of
technological changes (steam-powered presses and paper-making) and internal industrial developments
which lowered firms' cost of capital?  Or, rather, was the growth attributable to an expansion in
the number of novel readers or intensification of novel reading among the existing population of
novel readers? The latter, at least, seems unlikely, because the gains of the industrial
revolution---which might have enabled more people to purchase the luxury goods which novels and
circulating library subscriptions unquestionably were---did not accrue meaningfully to the broader
population until after \periodOfGrowthStart{} \parencite{allen2009engels}.

\section*{Conclusion}

The number of new novels published each year counts as essential information for researchers interested in understanding the text industry and text culture between 1789 and 1919.
Knowing that a novel was one among 100 (rather than 500) new works published in a given year affects how a researcher understands the position of a work in the literary marketplace \autocite{eliot2002very}.
Estimates of a variety of quantities which have been the subject of scholarly attention can be bounded by or estimated from the number of new novels published each year.
Novels' share of all editions can be bounded from below given the number of first edition novels and the number of works published in a given period.
(The changing share of prose fiction has been discussed in more than one scholarly study \autocite{erickson1996economy,eliot1998patterns}.)
A second quantity of interest to book historians and social historians of literature is the number of individuals who pursued careers as novelists \autocite{sutherland1988publishing}.
As the vast majority of novels are written by one person, this quantity can be bounded from above by the number of novels published during a given period.
Equipped with an estimate of the average number of novels published by a novelist during the period, a serviceable estimate of the quantity itself could be calculated as well.

Reliable estimates of the number of new novels published each year help bibliographers assembling exhaustive lists of published novels.
Such estimates allow bibliographers to gauge their progress.
For example, if a model such as ours, one which draws together a range of sources, predicts that there are very likely between 78 and 160 first edition women-authored novels published in 1865, a bibliographer can consult their list of titles to see if their total aligns with the estimate.
If the total in the bibliography falls conspicuously short of the estimated total, this indicates that novels by women are missing from the bibliography.
In such a scenario, the bibliographer might then expand the range of sources they are drawing on to identify novels.
Absent such estimates it is difficult for a bibliographer to conveniently assess their progress towards attaining an exhaustive list.
Without estimates of the total, they must follow the expensive and time-consuming approach of RFGS:
make sure they have exhaustively reviewed all sources of information that could have recorded the publication of a novel.
Equipped with good estimates of the total number of published titles, bibliographers can judge their progress at much lower cost.

Credible estimates of the share of new novels published during each year by gender allow literary studies scholars and book historians to assess how well arbitrary collections of novels reflects the population.
We have already mentioned a particular corpus, the ``reprint canon'', which includes novels widely-used in university teaching and research.
Our estimates allow us to compare the reprint canon to the population of published novels.
Another corpus of novels which might be compared with the population is the collection of novels authored by writers who are included in the Dictionary of National Bibliography (DNB).
If this corpus does not resemble the relevant population then it is unlikely that the individuals in the DNB resemble the population of novelists.
Knowing if the DNB reflects the population of novelists would permit researchers to calibrate their trust in existing studies which assume or suggest that writers in the DNB resemble the population (e.g., \textcite{altick1962sociology}).

The utility of the estimates presented here pales in comparison to the usefulness of an exhaustive bibliography of the \NewNovelsSeventeenHundredEightyNineToNineteenHundredNineteenFifthPercentile{}–\NewNovelsSeventeenHundredEightyNineToNineteenHundredNineteenNinetyFifthPercentile{} new novels published between 1789 and 1919.
The latter would allow us to say a great deal more about the particular kinds of novels which were published and the range of writers and publishers involved in the text industry.
But an exhaustive bibliography of new novels published in the British Isles and Ireland between 1789 and 1919 does not exist and is unlikely to emerge in the next few years.
In the interim, the estimates gathered here give researchers, bibliographers in particular, a series of bearings which will allow them to better assess existing accounts of the history of the novel and the history of the text industry.

\singlespacing

\printbibliography

\newpage

\section*{Appendix}

\begin{longtable}{lllll}
\toprule
{} & Novels, Men-authored & Novels, Women-authored & Novels, Unknown &    Novels, All \\
\midrule
\endhead
\midrule
\multicolumn{5}{r}{{Continued on next page}} \\
\midrule
\endfoot

\bottomrule
\endlastfoot
1789 &                 4-28 &                  20-50 &            8-35 &             71 \\
1790 &                 5-29 &                  17-49 &           12-41 &             74 \\
1791 &                10-39 &                  14-44 &            9-37 &             74 \\
1792 &                10-34 &                  20-44 &            0-12 &             58 \\
1793 &                 6-24 &                  13-33 &            2-17 &             45 \\
1794 &                12-36 &                  10-34 &            2-21 &             56 \\
1795 &                11-33 &                  14-36 &            0-10 &             50 \\
1796 &                17-53 &                  27-64 &            2-26 &             91 \\
1797 &                 2-25 &                  27-60 &            9-39 &             79 \\
1798 &                 7-35 &                  30-61 &            1-22 &             75 \\
1799 &                10-45 &                  46-84 &            0-20 &             99 \\
1800 &                   28 &                     41 &              13 &             82 \\
1801 &                   18 &                     44 &              12 &             74 \\
1802 &                   22 &                     29 &              10 &             61 \\
1803 &                   35 &                     33 &              11 &             79 \\
1804 &                   37 &                     30 &               8 &             75 \\
1805 &                   30 &                     34 &              12 &             76 \\
1806 &                   23 &                     39 &              10 &             72 \\
1807 &                   31 &                     30 &               8 &             69 \\
1808 &                   42 &                     49 &              20 &            111 \\
1809 &                   32 &                     37 &              10 &             79 \\
1810 &                   22 &                     51 &              18 &             91 \\
1811 &                   32 &                     36 &              12 &             80 \\
1812 &                   17 &                     33 &              17 &             67 \\
1813 &                   19 &                     39 &               6 &             64 \\
1814 &                   17 &                     41 &               5 &             63 \\
1815 &                   19 &                     23 &              12 &             54 \\
1816 &                   16 &                     30 &              13 &             59 \\
1817 &                   16 &                     30 &               9 &             55 \\
1818 &                   19 &                     31 &              13 &             63 \\
1819 &                   20 &                     32 &              21 &             73 \\
1820 &                   34 &                     28 &               8 &             70 \\
1821 &                   35 &                     27 &              13 &             75 \\
1822 &                   39 &                     29 &              14 &             82 \\
1823 &                   43 &                     32 &              13 &             88 \\
1824 &                   47 &                     40 &              12 &             99 \\
1825 &                   55 &                     25 &              12 &             92 \\
1826 &                   42 &                     28 &               7 &             77 \\
1827 &                   40 &                     29 &              11 &             80 \\
1828 &                   43 &                     27 &              13 &             83 \\
1829 &                   48 &                     24 &              11 &             83 \\
1830 &                33-69 &                  18-44 &            6-22 &            108 \\
1831 &                29-62 &                  15-40 &            6-20 &             69 \\
1832 &                31-66 &                  16-44 &            6-21 &             88 \\
1833 &                29-64 &                  15-44 &            6-20 &             80 \\
1834 &                28-66 &                  15-46 &            5-21 &             77 \\
1835 &                31-74 &                  17-53 &            6-24 &            112 \\
1836 &                28-69 &                  16-50 &            6-23 &             78 \\
1837 &                28-73 &                  16-56 &            6-24 &         71-128 \\
1838 &                28-74 &                  17-58 &            6-25 &         72-131 \\
1839 &                28-79 &                  18-63 &            7-27 &         77-138 \\
1840 &                27-80 &                  18-65 &            7-28 &         74-144 \\
1841 &                28-80 &                  18-67 &            7-29 &         79-144 \\
1842 &                29-83 &                  20-71 &            8-31 &         83-152 \\
1843 &                28-86 &                  21-73 &            8-32 &         84-156 \\
1844 &                29-88 &                  22-77 &            9-34 &         89-163 \\
1845 &                29-88 &                  23-79 &            9-35 &         90-167 \\
1846 &                31-95 &                  25-86 &           10-39 &         96-179 \\
1847 &               34-104 &                  29-96 &           11-43 &        108-203 \\
1848 &               36-110 &                 31-103 &           13-46 &        115-215 \\
1849 &               37-113 &                 34-107 &           13-48 &        118-226 \\
1850 &               41-124 &                 39-117 &           15-54 &        131-249 \\
1851 &               45-129 &                 42-124 &           16-57 &        138-264 \\
1852 &               46-131 &                 44-126 &           17-58 &        143-269 \\
1853 &               48-134 &                 47-129 &           18-60 &        147-276 \\
1854 &               51-136 &                 51-135 &           19-62 &        154-289 \\
1855 &               51-132 &                 51-132 &           19-60 &        151-284 \\
1856 &               50-127 &                 52-127 &           19-58 &        147-277 \\
1857 &               52-129 &                 53-130 &           19-58 &        150-283 \\
1858 &               54-127 &                 56-131 &           19-59 &        151-286 \\
1859 &               57-131 &                 59-135 &           20-59 &        156-297 \\
1860 &               60-131 &                 64-137 &           22-59 &        165-303 \\
1861 &               61-134 &                 64-140 &           20-58 &        165-308 \\
1862 &               62-135 &                 65-141 &           20-56 &        163-310 \\
1863 &               62-135 &                 66-141 &           20-55 &        163-310 \\
1864 &               65-142 &                 70-150 &           20-56 &        172-325 \\
1865 &               73-149 &                 78-160 &           23-58 &        190-350 \\
1866 &               71-152 &                 76-160 &           21-55 &        184-344 \\
1867 &               73-156 &                 78-165 &           20-54 &        188-353 \\
1868 &               77-165 &                 83-175 &           20-55 &        197-370 \\
1869 &               78-164 &                 84-177 &           20-53 &        199-371 \\
1870 &               82-167 &                 85-174 &           19-48 &        201-369 \\
1871 &               83-179 &                 93-197 &           20-53 &        216-404 \\
1872 &               84-179 &                 95-198 &           20-52 &        217-408 \\
1873 &               88-187 &                102-210 &           20-52 &        226-423 \\
1874 &               86-183 &                101-209 &           19-50 &        223-420 \\
1875 &               93-188 &                115-225 &           20-51 &        245-441 \\
1876 &               97-203 &                117-237 &           20-52 &        252-470 \\
1877 &              103-216 &                124-251 &           20-54 &        264-493 \\
1878 &              109-225 &                131-263 &           20-55 &        280-519 \\
1879 &              118-243 &                141-284 &           21-57 &        302-556 \\
1880 &              125-244 &                146-282 &           22-56 &        312-559 \\
1881 &              125-257 &                143-287 &           20-56 &        310-570 \\
1882 &              126-260 &                141-285 &           20-54 &        312-569 \\
1883 &              145-294 &                159-315 &           22-60 &        351-635 \\
1884 &              156-312 &                162-326 &           22-61 &        366-663 \\
1885 &              151-288 &                157-296 &           21-54 &        349-617 \\
1886 &              156-305 &                152-299 &           21-56 &        350-629 \\
1887 &              171-333 &                159-316 &           22-60 &        377-676 \\
1888 &              192-374 &                172-339 &           25-65 &        415-740 \\
1889 &              193-376 &                167-327 &           24-64 &        410-733 \\
1890 &              211-394 &                172-326 &           25-63 &        435-753 \\
1891 &              206-378 &                165-313 &           25-63 &        422-721 \\
1892 &              223-428 &                174-345 &           27-70 &        450-813 \\
1893 &              234-447 &                178-352 &           29-74 &        469-834 \\
1894 &              249-466 &                183-357 &           31-76 &        492-860 \\
1895 &              274-504 &                201-378 &           35-81 &        537-934 \\
1896 &              266-510 &                185-374 &           34-84 &        515-931 \\
1897 &              300-581 &                204-412 &           39-96 &      576-1,047 \\
1898 &              302-592 &                201-414 &           39-99 &      571-1,053 \\
1899 &              309-618 &                202-425 &          41-106 &      589-1,091 \\
1900 &              314-595 &                202-397 &          44-104 &      590-1,056 \\
1901 &              292-599 &                183-409 &          40-108 &      557-1,050 \\
1902 &              326-696 &                200-469 &          45-128 &      629-1,215 \\
1903 &              351-776 &                212-523 &          49-148 &      687-1,352 \\
1904 &              356-815 &                210-548 &          50-161 &      704-1,412 \\
1905 &              359-854 &                208-579 &          51-172 &      717-1,459 \\
1906 &              372-920 &                210-629 &          53-192 &      757-1,581 \\
1907 &            397-1,046 &                222-713 &          57-223 &      828-1,760 \\
1908 &            401-1,099 &                219-766 &          58-243 &      839-1,847 \\
1909 &            413-1,200 &                223-835 &          61-275 &      893-2,011 \\
1910 &            414-1,258 &                221-880 &          60-292 &      907-2,107 \\
1911 &            414-1,310 &                214-923 &          60-314 &      926-2,188 \\
1912 &            430-1,406 &              219-1,000 &          60-349 &      964-2,356 \\
1913 &            419-1,483 &              216-1,048 &          60-375 &      968-2,491 \\
1914 &            399-1,466 &              202-1,038 &          55-378 &      913-2,446 \\
1915 &            374-1,427 &              187-1,017 &          50-376 &      872-2,396 \\
1916 &            334-1,322 &                167-963 &          45-361 &      801-2,228 \\
1917 &            306-1,267 &                152-925 &          40-345 &      738-2,142 \\
1918 &            287-1,251 &                142-898 &          36-349 &      703-2,100 \\
1919 &            293-1,327 &                145-980 &          37-382 &      731-2,238 \\
All  &        18,344-33,508 &          14,395-26,402 &     3,720-8,050 &  39,631-62,978 \\

\caption{\textbf{New novels published between 1789 and 1919.} Intervals show 90\% credible
        intervals. Percentages are calculated with respect to table rows. Where intervals do not appear (1789--1836),
        counts shown are from RFGS. RFGS provide counts of new novels by author gender for 1800-1829 and total new novels for all years between 1789 and 1836.\label{tbl:novels-by-year}}

\end{longtable}

\begin{longtable}{lrrr}
\toprule
Gender &  Men-authored &  Women-authored &  Unknown \\
Year &               &                 &          \\
\midrule
\endhead
\midrule
\multicolumn{4}{r}{{Continued on next page}} \\
\midrule
\endfoot

\bottomrule
\endlastfoot
1789 &             0 &               0 &        0 \\
1790 &             0 &               2 &        0 \\
1791 &             0 &               3 &        0 \\
1792 &             0 &               1 &        0 \\
1793 &             0 &               1 &        0 \\
1794 &             1 &               1 &        0 \\
1795 &             0 &               1 &        0 \\
1796 &             2 &               4 &        0 \\
1797 &             0 &               2 &        0 \\
1798 &             0 &               1 &        0 \\
1799 &             2 &               1 &        0 \\
1800 &             1 &               2 &        0 \\
1801 &             2 &               3 &        0 \\
1802 &             0 &               0 &        0 \\
1803 &             0 &               0 &        0 \\
1804 &             0 &               0 &        0 \\
1805 &             1 &               1 &        0 \\
1806 &             0 &               2 &        0 \\
1807 &             0 &               0 &        0 \\
1808 &             0 &               1 &        1 \\
1809 &             0 &               1 &        0 \\
1810 &             1 &               1 &        0 \\
1811 &             0 &               2 &        0 \\
1812 &             0 &               1 &        0 \\
1813 &             0 &               1 &        0 \\
1814 &             1 &               1 &        0 \\
1815 &             2 &               0 &        0 \\
1816 &             1 &               1 &        0 \\
1817 &             2 &               1 &        0 \\
1818 &             2 &               3 &        0 \\
1819 &             1 &               0 &        0 \\
1820 &             2 &               0 &        0 \\
1821 &             1 &               0 &        0 \\
1822 &             0 &               0 &        0 \\
1823 &             0 &               1 &        0 \\
1824 &             2 &               0 &        0 \\
1825 &             0 &               0 &        0 \\
1826 &             0 &               1 &        0 \\
1827 &             1 &               1 &        0 \\
1828 &             0 &               0 &        0 \\
1829 &             0 &               0 &        0 \\
1830 &             0 &               0 &        0 \\
1831 &             0 &               0 &        0 \\
1832 &             0 &               0 &        0 \\
1833 &             0 &               0 &        0 \\
1834 &             0 &               0 &        0 \\
1835 &             0 &               1 &        0 \\
1836 &             1 &               0 &        0 \\
1837 &             1 &               0 &        0 \\
1838 &             1 &               0 &        0 \\
1839 &             2 &               0 &        0 \\
1840 &             2 &               0 &        0 \\
1841 &             3 &               0 &        0 \\
1842 &             0 &               0 &        0 \\
1843 &             1 &               0 &        0 \\
1844 &             3 &               0 &        0 \\
1845 &             1 &               0 &        0 \\
1846 &             2 &               0 &        0 \\
1847 &             0 &               3 &        0 \\
1848 &             3 &               2 &        0 \\
1849 &             1 &               1 &        0 \\
1850 &             1 &               0 &        0 \\
1851 &             1 &               0 &        0 \\
1852 &             2 &               0 &        0 \\
1853 &             2 &               3 &        0 \\
1854 &             2 &               0 &        0 \\
1855 &             1 &               1 &        0 \\
1856 &             0 &               1 &        0 \\
1857 &             4 &               1 &        0 \\
1858 &             1 &               1 &        0 \\
1859 &             1 &               1 &        0 \\
1860 &             2 &               1 &        0 \\
1861 &             2 &               2 &        0 \\
1862 &             3 &               1 &        0 \\
1863 &             2 &               3 &        0 \\
1864 &             3 &               1 &        0 \\
1865 &             2 &               2 &        0 \\
1866 &             1 &               3 &        0 \\
1867 &             4 &               1 &        0 \\
1868 &             1 &               0 &        0 \\
1869 &             3 &               0 &        0 \\
1870 &             2 &               0 &        0 \\
1871 &             4 &               1 &        0 \\
1872 &             5 &               1 &        0 \\
1873 &             3 &               1 &        0 \\
1874 &             3 &               0 &        0 \\
1875 &             3 &               0 &        0 \\
1876 &             2 &               2 &        0 \\
1877 &             2 &               1 &        0 \\
1878 &             2 &               0 &        0 \\
1879 &             4 &               0 &        0 \\
1880 &             2 &               2 &        0 \\
1881 &             4 &               0 &        0 \\
1882 &             2 &               0 &        0 \\
1883 &             3 &               2 &        0 \\
1884 &             1 &               0 &        0 \\
1885 &             1 &               0 &        0 \\
1886 &             6 &               0 &        0 \\
1887 &             2 &               1 &        0 \\
1888 &             6 &               2 &        0 \\
1889 &             5 &               0 &        0 \\
1890 &             3 &               2 &        0 \\
1891 &             5 &               0 &        0 \\
1892 &             3 &               0 &        0 \\
1893 &             2 &               0 &        0 \\
1894 &             4 &               3 &        0 \\
1895 &             5 &               0 &        0 \\
1896 &             2 &               0 &        0 \\
1897 &            10 &               1 &        0 \\
1898 &             2 &               0 &        0 \\
1899 &             3 &               0 &        0 \\
1900 &             2 &               0 &        0 \\
1901 &             4 &               1 &        0 \\
1902 &             5 &               1 &        0 \\
1903 &             5 &               0 &        0 \\
1904 &             2 &               0 &        0 \\
1905 &             3 &               0 &        0 \\
1906 &             3 &               0 &        0 \\
1907 &             2 &               0 &        0 \\
1908 &             6 &               0 &        0 \\
1909 &             2 &               0 &        0 \\
1910 &             3 &               0 &        0 \\
1911 &             3 &               2 &        0 \\
1912 &             2 &               0 &        0 \\
1913 &             2 &               0 &        0 \\
1914 &             2 &               0 &        0 \\
1915 &             0 &               0 &        0 \\
1916 &             0 &               0 &        0 \\
1917 &             0 &               0 &        0 \\
1918 &             0 &               0 &        0 \\
1919 &             0 &               0 &        0 \\

\caption{\textbf{Novels published between 1789 and 1919 which are still in print.} The table shows counts of novels originally published between 1789 and 1919 available from Broadview Press, Penguin, or Oxford in 2018. Sources: Broadview Press 2018 English Catalogue, Penguin Classics 2016 Catalog, Oxford World's Classics 2016 Catalog.\label{tbl:novels-reprint-canon-by-year}}

\end{longtable}

\subsection*{Source code and datasets}

Source code and data used accompany this paper.

\section*{Acknowledgments}

We thank participants, in particular Fotis Jannidis and Karina van Dalen-Oskam, in the Symposium ``Digitale
Literaturwissenschaft'' for helpful comments.  Many thanks to Troy Bassett for his willingness to
participate in the (lengthly) elicitation task and for his advice on collecting the reprint canon
data. We are grateful to Laura Schneider for excellent research assistance.

\end{document}